\documentclass[a4paper,11pt]{article}
\usepackage{jheppub} 
\usepackage{lineno}
\newcommand{\bea}{\begin{eqnarray}}
\newcommand{\beq}{\begin{equation}}

\newcommand{\eea}{\end{eqnarray}}
\newcommand{\eeq}{\end{equation}}
\newcommand{\ds}{\displaystyle}

\usepackage{amsmath}
\usepackage{amsfonts}
\usepackage{amssymb}
\usepackage{bm}       
\usepackage[usenames,dvipsnames]{color}
\usepackage{comment}
\usepackage{dcolumn}  
\usepackage{epsfig}
\usepackage{graphics}
\usepackage{graphicx} 
\usepackage{lipsum}
\usepackage[rightcaption]{sidecap}
\usepackage{soul}

\usepackage{placeins} 
\usepackage{psfrag}
\usepackage{times}
\usepackage[varg]{txfonts}
\usepackage{xcolor}
\usepackage{xspace}
\graphicspath{%
    {converted_graphics/}
    {/}
}

\begin{document}

\title{Dispersion relation, propagation features and
canonical gauge structure of GNLED  }

\author[a,b,c]{Abedennour Dib }
\emailAdd{ abedennour.dib@cnrs-orleans.fr}
\affiliation[a]{Laboratoire de Physique et Chimie de l'Environnement et de l'Espace (LPC2E) UMR 7328, Centre National de la Recherche Scientifique (CNRS), Universit\'e d'Orl\'eans (UO) Centre National d'\'Etudes Spatiales (CNES), 3A Avenue de la Recherche Scientifique, 45071 Orl\'eans, France
}
\affiliation[b]{Observatoire des Sciences de l'Univers en region Centre (OSUC) UMS 3116
 Universit\'e d'Orl\'eans (UO), Centre National de la Recherche Scientifique (CNRS), Observatoire de Paris (OP), Universit\'e Paris Sciences \& Lettres (PSL), 1A rue de la F\'{e}rollerie, 45071 Orl\'{e}ans, France}
\affiliation[c]{D\'epartement de Physique, Unit\'e de Formation et Recherche Sciences et Techniques, Universit\'e d'Orl\'eans Rue de Chartres, 45100 Orl\'{e}ans, France}
\author[d]{Jos\'e A. Helay\"el-Neto }
\emailAdd{ helayel@cbpf.br}

\affiliation[d]{Departamento de Astrof\'{\i}sica, Cosmologia e Intera\c{c}\~{o}es Fundamentais (COSMO), Centro Brasileiro de Pesquisas F\'{\i}sicas (CBPF)
Rua Xavier Sigaud 150, 22290-180 Urca, Rio de Janeiro, RJ, Brazil}
\author[f]{Tomislav Prokopec }
\emailAdd{t.prokopec@uu.nl}
\affiliation[f]{Institute for Theoretical Physics, Spinoza Institute \& EMME $\Phi$, Utrecht University,
Postbus 80.195, 3508 TD Utrecht, The Netherlands}
\author[a,b,c]{Alessandro D.A.M. Spallicci  
}
\emailAdd{ spallicci@cnrs-orleans.fr}
\author[e]{Youssef Temmam }
\emailAdd{youssef.temmam@etu.univ-amu.fr}
\affiliation[e]{Aix-Marseille Université, 163 avenue de Luminy, 13288 Marseille, France}

\abstract{
 We investigate the propagation properties and canonical gauge structure of Generalised Non-Linear Electrodynamics (GNLED) in the presence of a non-dynamical background field. Starting from the quadratic photon sector of the theory, we perform a field redefinition that converts the background-induced derivative coupling into a mass-like contribution characterized by the vector \(V_\mu=\partial_\mu\alpha/\alpha\) where $\alpha= \sqrt{\left.\frac{\partial\mathcal{L}}{\partial \mathcal{F}}\right|_B}$. For a constant background, we derive the corresponding dispersion relation and show that, for a purely space-like \(V_\mu\), the propagating mode obeys \(\omega^2=\|\vec k\|^2+\|\vec V\|^2\), defining an effective mass scale \(m_{\rm eff}=|\vec{V}|\) and a finite rest frequency. The resulting phase and group velocities exhibit the characteristic dispersive behaviour of a gapped mode, with subluminal group velocity and superluminal phase velocity. We then construct the propagator and show that the massive pole resides in the transverse sector, while the longitudinal singularity is associated with gauge fixing. A complete Hamiltonian analysis yields two first-class constraints and only two physical degrees of freedom, demonstrating that the effective mass does not introduce an additional polarisation. The associated gauge transformation is a deformed Abelian \(U(1)\) symmetry. Finally, we show that the Hamiltonian is positive definite for purely space-like backgrounds.}

\date{\today} 


\maketitle

\section{ Introduction}

Classical electrodynamics (CED) is one of the cornerstones of modern physics, combining in a single coherent framework Maxwell's theory of electromagnetism with special relativity. The theory relies on four key features: gauge invariance which is linked to the masslessness of the photon, the linearity of Maxwell's equations, duality invariance in a vacuum and invariance under Lorentz symmetry. The quantization of classical electrodynamics and its interaction with charged matter gave rise to quantum electrodynamics (QED), beginning with the pioneering works of Dirac, Jordan, Pauli and Heisenberg \cite{Dirac-1927, Jordan-Pauli-1928, Heisenberg-Pauli-1929}. Building upon it, Feynman and Schwinger brought up a consistent renormalized theory of QED \cite{Schwinger-1948,Feynman-1949}. Nevertheless, quantum electrodynamics sacrifices one of the ingredients of classical electrodynamics as it introduces non-linear corrections to the theory.

 Particularly, non-linear electromagnetic effects emerge as corrections to the leading order linear theory. Such effects were first investigated by Sauter \cite{Sauter-1931}. Subsequently, Born and Infeld proposed a non-linear extension of Maxwell's electrodynamics, with the aim of resolving the divergence of the electron self-energy at the classical level \cite{born-infeld-1934a,born-infeld-1934b,born-infeld-1935}. De Broglie and Proca proposed an extension of electrodynamics in which the photon is endowed with a non-zero mass \cite{debroglie-1923,debroglie-1936,proca-1936d,proca-1937}. From a theoretical point of view, massive photons remain admissible provided that the effects associated with the additional longitudinal polarisation are sufficiently small to be undetected \cite{Yan:2023kdg,Novas-etal-2024,Ignatiev-Yoshi-1996}. A massive photon may also provide a natural infrared regulator for the divergences associated with soft photons \cite{Weinberg-1965}. Moreover, the exact masslessness of the photon cannot be established with absolute experimental certainty, in particular due to the fundamental limitations imposed by Heisenberg’s principle \cite{spallicci-benetti-capozziello-2022,Capozziello-Sarracino-Spallicci-2023}. Nevertheless, increasingly stringent upper bounds on the photon mass have been obtained from laboratory, astrophysical, and cosmological observations \cite{spallicci-etal-2024b}. 

Massive photon modes also arise in extensions of Maxwell electrodynamics. In particular, the Bopp-Podolsky model introduces, in addition to the usual massless electromagnetic mode, an extra massive mode \cite{bopp-1940}. The hypothesis of an effective photon mass is therefore not merely formal: massive electromagnetic modes also emerge naturally in condensed-matter systems, where they are essential to describe phenomena such as the Meissner effect \cite{Weinberg-1986,helayelneto-spallicci-2019,spallicci-etal-2021}.

Approaches combining both types of extensions have also been explored in the literature. More recently, considerable attention has been devoted to the class of Generalised Proca theories \cite{Heisenberg-2014}, including the recently proposed Proca--Nuevo model \cite{derham-etal-2023}. These frameworks are particularly interesting because they establish connections with theories of massive gravity \cite{derham-etal-2022,defelice-etal-2016}, another active area of research in theoretical physics. The work presented here falls within this broader class of extensions.

We showed previously that within the framework of Generalised Nonlinear Electrodynamics (GNLED), an effective massive mode can emerge \cite{Dib:2025qoy}. Contrary to what might be expected at first sight, the theory can nevertheless be formulated as a well defined gauge theory, given that the appropriate gauge transformations are employed. We will further show that, as a consequence, the corresponding massive polarisation must be accommodated within the transverse sector of the propagator, and we will discuss in greater detail the resulting gauge structure of the theory.

In this paper, we work in Minkowski spacetime, where the metric has as the convention $\eta^{\mu\nu}=\left(+,-,-,-\right)$. Spacetime indices are labelled with Greek symbols, whereas the time index as $0$ component and space indices with Latin symbols.

\section{Generalised Non-Linear Electrodynamics}
\subsection{Construction of the action}

GNLED is a family of electromagnetic models that are constructed as a power expansion of Maxwell invariants. In \cite{spallicci-etal-2024a}, we argue that these models are general in the sense that well-known non-linear extensions of electrodynamics are already encompassed within the framework namely Born-Infeld \cite{born-infeld-1934a,born-infeld-1934b,born-infeld-1935} and Heisenberg-Euler \cite{Heisenberg-Euler-1936}. Discussing the construction of the theory goes beyond the scope of this paper, and we will only offer here a brief summary. The two Maxwell invariants are defined as  
\begin{align}
   & \mathcal{F}=-\frac{1}{4\mu_0}F^{\mu\nu}F_{\mu\nu}=\frac{1}{2\mu_0}\left(\frac{\Vec{E}^2}{c^2}-\Vec{B}^2\right)~,  &
   \\
  & \mathcal{G}=-\frac{1}{4\mu_0}F^{\mu\nu}\Tilde{F}_{\mu\nu}=-\frac{1}{4\mu_0}F^{\mu\nu}\epsilon_{\mu\nu\rho\sigma}F^{\rho\sigma}=\frac{1}{\mu_0 c}\Vec{E}\cdot\Vec{B} ~,&
\end{align}
where $\Vec{E}$ is the electric field, $\Vec{B}$ the magnetic field and $F^{\mu\nu}$ the field strength tensor.

It is then possible to construct an action that is expressed as a power series expansion of said invariants up to arbitrary order such as
\begin{align}
        \mathcal{L}(\mathcal{F},\mathcal{G})=&\mathcal{L}_0+\frac{\partial \mathcal{L}}{\partial \mathcal{F}}\mathcal{F}+\frac{\partial \mathcal{L}}{\partial \mathcal{G}}\mathcal{G}+ \frac12 \frac{\partial^2 \mathcal{L}}{\partial {\mathcal{F}}^2}\mathcal{F}^2+\frac12 \frac{\partial^2 \mathcal{L}}{\partial {\mathcal{G}}^2}\mathcal{G}^2+ \frac{\partial^2 \mathcal{L}}{\partial \mathcal{F}\partial \mathcal{G}}\mathcal{F}\mathcal{G}+...&~.
\end{align}

This family of models in its initial formulation falls under the Plebanski class of electrodynamics: models that are both Lorentz invariant and gauge invariant under U(1) transformations \cite{Plebankski-1970, Schellestede-Perlick-Lammerzahl-2016}. Massive photons are incompatible with the Plebanski class of electrodynamics \cite{Plebankski-1970}, and we wish to modify the theory enough to leave it, without totally breaking down the structure of the model. To achieve this, we will split the electromagnetic field into a background and a photon field that acts as a perturbation. In the language of effective field theories, we impose that the background field has an infrared (IR) behaviour while the perturbation (which we will refer to as photon) encompasses the ultraviolet (UV) character of the model, disconnecting the two entities but not forbidding their interactions. We will refer to their field strength and potentials with upper and lower case characters, respectively for background and photon. With this new notation, we
start by defining the total field strength tensor as
\begin{align}
    F^{\mu\nu}_{Total}={F}^{\mu\nu}+f^{\mu\nu} ~,
\end{align}
where $F^{\mu\nu}=\partial_\mu A_\nu-\partial_\nu A_\mu$ and $f^{\mu\nu}=\partial_\mu a_\nu-\partial_\nu a_\mu$ are respectively the background and photon field-strength tensors. For the Maxwell invariants, it gives 
\begin{align}
   & \mathcal{F}_{Total}= {F}^{\mu\nu}{F}_{\mu\nu}+2 {F}^{\mu\nu}f_{\mu\nu}+ f^{\mu\nu}f_{\mu\nu}= \mathcal{F}+\delta \mathcal{F} ~,& \\
   & \mathcal{G}_{Total}= \tilde{F}^{\mu\nu}{F}_{\mu\nu}+2 \tilde{F}^{\mu\nu}f_{\mu\nu}+\tilde{f}^{\mu\nu}f_{\mu\nu}= \mathcal{G}+\delta \mathcal{G}~. & 
\end{align}

 Before going further, we will move to the natural unit system where $h=c=\mu_0=1$. Using the power series expansion around the fixed background would result in the following Lagrangian density at quadratic order in the photon field-strength $f^{\mu\nu}$ would then be \cite{spallicci-etal-2024a, Dib:2025qoy} 
\begin{align}
    &\mathcal{L}(\mathcal{F},\mathcal{G})_{{\rm Total}}=\mathcal{L}(\mathcal{F}+\delta \mathcal{F}, \mathcal{G}+\delta \mathcal{G}) \\
    &=\mathcal{L}(\mathcal{F},\mathcal{G})+\sum_{n=1}^i\frac{1}{n!}\left(\delta \mathcal{F}\frac{\partial}{\partial\mathcal{F}}+\delta \mathcal{G}\frac{\partial}{\partial \mathcal{G}}\right)^n\mathcal{L} \Big|_B \nonumber \\
    & \approx \mathcal{L}(\mathcal{F},\mathcal{G})+\delta \mathcal{F}\left.\frac{\partial\mathcal{L}}{\partial \mathcal{F}}\right\vert_B+\delta \mathcal{G}\left.\frac{\partial\mathcal{L}}{\partial \mathcal{G}}\right\vert_B  +\frac12\left(\delta \mathcal{F}\right)^2\left.\frac{\partial^2\mathcal{L}}{\partial \mathcal{F}^2} \right\vert_B
    +\frac12\left(\delta \mathcal{G}\right)\left.\frac{\partial^2\mathcal{L}}{\partial \mathcal{G}^2}\right\vert_B + \delta \mathcal{F}\delta \mathcal{G}\left.\frac{\partial^2\mathcal{L}}{\partial \mathcal{F}\partial \mathcal{G}}\right\vert_B \\
    &=\mathcal{L}_{B}+ \mathcal{L}_{2} ~, &
\end{align}
where $\mathcal{L}_{B}$ represents the entirety of the pure background contributions and  $\mathcal{L}_2$ contains the photon terms and photon-background interactions of at most up to the second order in $f^{\mu\nu}$ . It is assumed that said expansion is well defined, {\it{i.e.}} that it converges and that each order contributes less than the previous one. We also define the coefficients 
\begin{align}
    &C_1=\left.\frac{\partial\mathcal{L}}{\partial \mathcal{F}}\right\vert_B~, &C_2=\left.\frac{\partial\mathcal{L}}{\partial \mathcal{G}}\right\vert_B ~,&&D_1=\left.\frac{\partial^2\mathcal{L}}{\partial \mathcal{F}^2}\right\vert_B \nonumber ~,&\\ 
    &\left.D_3=\frac{\partial^2\mathcal{L}}{\partial \mathcal{G}^2}\right\vert_B ~,&\left.D_2=\frac{\partial^2\mathcal{L}}{\partial \mathcal{F}\partial \mathcal{G}}\right\vert_B~,&\nonumber
\end{align}

 Since we suppose that the background field is an entity that has only infrared behaviour, and far removed from the observer, we will consider it to be non-dynamical but nonetheless allow it to be non-constant in the most general case. With the previous assumptions, we will ignore the $\mathcal{L}_B$ term and only consider the  $\mathcal{L}_2$ contributions which will read
\begin{align}\label{actionorder2}
    \mathcal{L}_2=&-\frac12\left(C_1{F}^{\mu\nu}+C_2{\tilde{F}}^{\mu\nu}\right)f_{\mu\nu}-\frac14 C_1f^{\mu\nu}f_{\mu\nu}-\frac14 C_2 f^{\mu\nu}\tilde{f}_{\mu\nu}+\frac18 Q^{\mu\nu\rho\sigma}f_{\mu\nu}f_{\rho\sigma}~,
\end{align}
where $Q$ is a shorthand notation for 
\begin{align}
    &{Q}^{[\mu\nu][\rho\sigma]}=D_1{F}^{\mu\nu}{F}^{\rho\sigma}+D_3 {\tilde{F}}^{\mu\nu}{\tilde{F}}^{\rho\sigma}+2D_2{F}^{\mu\nu}{\tilde{F}}^{\rho\sigma}~~,
\end{align}
the indices $\{\mu,\nu\}$ and $\{\rho,\sigma\} $ are antisymmetric within their respective brackets, while $\left[\mu\nu, \rho\sigma\right]$ are symmetric between themselves; finally,  $\{\mu,\rho\}, \{\mu,\sigma\}, \{\nu,\rho\},\{\nu,\sigma\}$ do not commute.\\

The variation of the action $\delta \mathcal{L}_2 $ with respect to the photon four potential $a_\mu$, yields the field equations 
\begin{align}
     &-\partial_\mu\left(C_1 {F}^{\mu\nu}+C_2{\tilde{F}}^{\mu\nu}\right)-\partial_\mu \left(C_1 f^{\mu\nu}+C_2 \tilde{f}^{\mu\nu}\right)+\frac12\partial_\mu\left({Q}^{\mu\nu\rho\sigma}f_{\rho\sigma}\right)=0~,&
\end{align}
which are the modified Maxwell equations for the theory \cite{spallicci-etal-2024a, Dib:2025qoy} in the general case of non-constant but non-dynamical backgrounds. If said background is taken to be a constant, the equations of motion simplify to 
\begin{align}
   - C_1\partial_\mu f^{\mu\nu}-C_2 \partial_\mu \tilde{f}^{\mu\nu}+\frac12Q^{\mu\nu\rho\sigma}f_{\rho\sigma}\partial_\mu f_{\rho\sigma}=0~.
\end{align}

\subsection{Field redefinition}
Before going any further, we will highlight that, if we were to be strict, the action is not exactly quadratic in $f^{\mu\nu}$. Indeed, considering \eqref{actionorder2} 
\begin{align}
    \mathcal{L}_2=&-\frac12\left(C_1{F}^{\mu\nu}+C_2{\tilde{F}}^{\mu\nu}\right)f_{\mu\nu}-\frac14 C_1f^{\mu\nu}f_{\mu\nu}-\frac14 C_2 f^{\mu\nu}\tilde{f}_{\mu\nu}+\frac18 Q^{\mu\nu\rho\sigma}f_{\mu\nu}f_{\rho\sigma}~,
\end{align}
we can see that the first term is actually linear in $f$, and thus must act as a current. Furthermore, while the remaining terms are quadratic in $f$, they exhibit different orders in background fields. Since the coefficients $C_1$ and $C_2$ are purely background dependent, but dimensionless, we will refer to them as $\mathcal{O}(0)$ contributions. The coefficients $D_1, D_2$ and $D_3$ will be of order $\mathcal{O}(1)$ in background. From this, and the supposed well defined nature of the power series expansion, we will consider that the $Q^{\mu\nu\rho\sigma}$ contributions are suppressed in comparison to the $C_1$ and $C_2$ ones, and neglect them in what follows.  

This reduces the action to the following expression when considering only background $\mathcal{O}(0)$ terms
\begin{align}
    \mathcal{L}_1= -\frac12\left(C_1{F}^{\mu\nu}+C_2{\tilde{F}}^{\mu\nu}\right)f_{\mu\nu}-\frac14 C_1f^{\mu\nu}f_{\mu\nu}
    &-\frac14 C_2 f^{\mu\nu}\tilde{f}_{\mu\nu}~,
\end{align}
which is labelled as $\mathcal{L}_1$ as it is equivalent to proceeding with the power series expansion up to first order in $\delta \mathcal{F}$ and $\delta \mathcal{G}$.

We will also drop dual contributions by setting $C_2=0$. This may seem motivated by physical reasons since we could naively assume that the $C_2$ terms are topological, but in our case $C_2$ being a variable makes said assumption incorrect. In our situation, it is a simplification choice to make our argument clearer but we will address some concerns that could arise later on.

This leads then to the following form 
\begin{align}
    \mathcal{L}_1=&-\frac14 C_1f^{\mu\nu}f_{\mu\nu}~-\frac12\left(C_1{F}^{\mu\nu}\right)f_{\mu\nu}~,
\end{align}
in which we can integrate by parts the second term and get
\begin{align}\label{ActionOrder1}
       \mathcal{L}_1= -\frac14 C_1 f^{\mu\nu}f_{\mu\nu} + a_\nu \mathcal{J}^\nu~,
\end{align}
where $\mathcal{J}^\nu= \partial_\mu \left(C_1 F^{\mu\nu}\right)$  will act as a conserved current if the background field is taken to be non-constant. Taking the variation of $\mathcal{L}_1$ with respect to the photon four potential $a^\mu$ gives us the equations of motion
\begin{align}
    \partial_\mu (C_1f^{\mu\nu})+\mathcal{J^\nu}=0~.
\end{align}
In the case where the background field is taken to be constant, the solution reduces to the usual Maxwell theory since
\begin{align}
    C_1 \partial_\mu f^{\mu\nu}=0~,
\end{align}
is totally independent from the background field.

However, if the background is taken to be non constant we get 
\begin{align}
    \partial_\mu C_1 f^{\mu\nu}+ C_1 \partial_\mu f^{\mu\nu}= -\mathcal{J}^\nu~,
\end{align}
which highlights an important fact: that the photon is never propagating in a vacuum, despite the absence of external sources due to the background contribution. This has already been discussed previously, and we could naïvely neglect this source term. However, we would still be left with
\begin{align}
    \partial_\mu f^{\mu\nu}= -\frac{1}{C_1} \partial_\mu C_1 f^{\mu\nu}~,
\end{align}
which infers that the photon is never freely propagating, as it seems fundamentally coupled with the background field. Furthermore, we can define a longitudinal four vector $C^\mu=\displaystyle\frac{\partial^\mu C_1}{C_1}$, which will then recast the equation of motion as 
\begin{align}
    (\partial_\mu+C_\mu)f^{\mu\nu}=0~.
\end{align}

This is a structural problem in the model, as it seems that we may not retrieve the usual dispersion relations in any limit as long as $C_1 \neq constant$, and that the four wave vector $k^\mu$ is not a null vector of the theory anymore, altering its gauge structure. To address both issues, we will redefine the potentials as done in \cite{Dib:2025qoy} and consider that    
\begin{align} \label{fieldredif}
    a'^\mu \rightarrow \alpha ~a^\mu~, 
\end{align} with $\alpha^2= C_1$. Constructing a field strength tensor for the redefined potential $a'^\mu$ gives
\begin{align}
f'^{\mu\nu}=\partial^\mu a'^\nu - \partial^\nu a'^\nu = \alpha f^{\mu\nu}+ a^\nu \partial^\mu \alpha- a^\mu \partial^\nu \alpha&~~,
\end{align}
and thus
\begin{align}
    \alpha f^{\mu\nu}=f'^{\mu\nu}-\frac{1}{\alpha}\left( a'^\nu \partial^\mu \alpha- a'^\mu \partial^\nu \alpha\right)~~.
\end{align}
By defining then $V^\mu = {\ds \frac{\partial^\mu \alpha}{\alpha}}$, we get the following expression
\begin{align}
    \alpha f^{\mu\nu}=f'^{\mu\nu}-\left( a'^\nu V^\mu- a'^\mu V^\nu\right)&~~,
\end{align}
which is breaking the manifest gauge invariance. Indeed, while the left hand side is trivially invariant by the usual $U(1)$ transformations, the right hand side is not. We will show later on that, unlike what was claimed naively in \cite{Dib:2025qoy}, gauge invariance is not lost, but that the set of transformations the theory obeys is, hence that the gauge symmetry is still respected when viewed in the appropriate variables. 

Proceeding further, we can square the field strength
\begin{align}
     \alpha^2 f^{\mu\nu}f_{\mu\nu}=f'^{\mu\nu}f'_{\mu\nu}+ 4 a'_\mu V_\nu f'^{\mu\nu}+2 \left(a'^\nu a'_\nu V^\mu V_\mu- a'_\mu a'_\nu V^\mu V^\nu\right)&~~,
\end{align}
and get the equivalent form of the action
\begin{align}\label{action 1}
    \mathcal{L}'_0= -\frac14f'^{\mu\nu}f'_{\mu\nu}-a'_\mu V_\nu f'^{\mu\nu}-\frac12 \left(a'^\nu a'_\nu V^\mu V_\mu- a'_\mu a'_\nu V^\mu V^\nu\right)~~,
\end{align}
where the last term will act as a mass, and can be equivalently written as 
\begin{align}
   -\frac12 \left(a'^\nu a'_\nu V^\mu V_\mu- a'_\mu a'_\nu V^\mu V^\nu\right) = \frac14(\varepsilon^{\mu\nu\rho\sigma}a_\rho V_\sigma)^2~,
\end{align}
we will however continue to use the original expression in what follows.

\subsection{Equations with of motion and dispersion relations}

In what follows, we will assume the background vector $V^\mu$ to be constant, in accordance with \cite{Dib:2025qoy}. However, we will refrain from imposing it to be purely space-like this early in the development. 

The equations of motion are obtained by taking the variation of the action with respect to the four potential $a'_\mu$  
\begin{align} \label{EOM'1}
    \partial_\mu f'^{\mu\nu}+ V^\nu \partial^\mu a'_\mu - V^\mu \partial^\nu a'_\mu-a'^{\nu} V^\mu V_\mu+ a'^\mu V^{\nu}V_\mu=0~~,
\end{align}
the divergence of \eqref{EOM'1} yields the following subsidiary condition
\begin{align}
    V_\nu \partial_\mu f'^{\mu\nu}+ V^2 \partial_\mu a'^\mu -V^\mu V^\nu \partial_\mu a'_\nu=0~~,
\end{align}
which as noted in \cite{Dib:2025qoy} hints towards the absence of gauge invariance. Indeed, the absence of subsidiary condition in Maxwell's theory is an expression of Noether's second theorem \cite{Avery:2015rga},  which indicates the presence of gauge transformations.

Having a closer look at the gauge invariance of the action under the usual gauge transformation for $U(1)$ theories $a''^\mu= a'^\mu + \partial^\mu \chi$  we get
\begin{align}
    \delta \mathcal{L}&= \delta\left[-\frac14f'^{\mu\nu}f'_{\mu\nu}-a'_\mu V_\nu f'^{\mu\nu}-\frac12 \left(a'^\nu a'_\nu V^\mu V_\mu- a'_\mu a'_\nu V^\mu V^\nu\right)\right] \nonumber\\
   & = \left(  V_\nu \partial_\mu f'^{\mu\nu}+ V^2 \partial_\mu a'^\mu -V^\mu V^\nu \partial_\mu a'_\nu\right) \chi~~,
\end{align}
which holds up to 4-divergences. This result confirms that the action is not gauge invariant. Indeed, even though the subsidiary condition is not independent from the equations of motion, gauge invariance has to upheld off-shell, confirming that the $U(1)$ transformations do not generate a Noether identity \cite{Avery:2015rga}. This is not surprising since $a'_\mu$ is not the original gauge field, but a transformed one. There is however still a gauge symmetry with respect to the original variable $a_\mu$ which is "hidden".

In \cite{Dib:2025qoy} we proceeded with the idea that the theory was not gauge invariant, and modified the action in a way that showed the presence of three degrees of freedom as well as a rest frame through the mechanical reduction. However, we believe that one of the modifications (which consisted on setting the totally symmetric and traceless part of the $V^\mu V^\nu$ tensor to zero) acted as an ill-suited gauge fixing, and altered the gauge structure of the theory. In what is next, we shall look at the most general case of the action. We will not set $V^\mu$ to be purely space-like, nor fix any parts of the action à priori. The only remaining condition, is that $V^\mu$ is taken to be constant.
\subsection{ Dispersion relations and refractive indices}
Consider the equations of motion for the theory
\begin{align} \label{EOM'1}
    \mathcal{E}^\nu=\partial_\mu f'^{\mu\nu}+ V^\nu \partial^\mu a'_\mu - V^\mu \partial^\nu a'_\mu-a'^{\nu} V^\mu V_\mu+ a'^\mu V^{\nu}V_\mu=0~~.
\end{align}
 By using the plane wave ansatz $\partial_\mu= -ik_\mu$, we get the Fourier space representation for the model
\begin{align}
    \left[-\eta^{\mu\nu}\left(k^2+V^2\right)+k^\mu k^\nu+iV^\mu k^\nu - iV^\nu k^\mu+ V^\mu V^\nu\right]\tilde{a}(k)_\mu=0~.
\end{align}
To obtain the dispersion relations, we must evaluate the determinant of the tensor
\begin{align}
D^{\mu\nu}=\left[-\eta^{\mu\nu}\left(k^2+V^2\right)+k^\mu k^\nu+iV^\mu k^\nu -i V^\nu k^\mu+ V^\mu V^\nu\right]~,
\end{align}
and then solve $\det(D^{\mu\nu})=0$.  We will start by rewriting $D^{\mu\nu}$ as
\begin{align}
    D^{\mu\nu}=-\zeta \eta^{\mu\nu}+ \chi^{\mu\nu}~,
\end{align}
where 
\begin{align}
&\zeta = k^2+V^2~, \\
   & \chi^{\mu\nu}=k^\mu k^\nu +iV^\mu k^\nu-i V^\nu k^\mu+ V^\mu V^\nu= \left(k^\mu + i V^\mu\right)\left(k^\nu-i V^\nu\right)~,
\end{align}
which then gives
\begin{align}\label{dispersionmatrix}
    D^{\mu\nu}=-\zeta \left(\eta^{\mu\nu}- \frac{1}{\zeta}\chi^{\mu\nu}\right)~,
\end{align}
where we will redefine the last term as $\xi^{\mu\nu}=\frac{1}{\zeta}\chi^{\mu\nu}$ and then express the full determinant as a binomial expansion
\begin{align}
    &\det(D)= \left(-\zeta\right)^4\det(\eta^{\mu\nu}-\xi^{\mu\nu})~,\\
    &\det( \eta-\xi)=1- {\xi^{\mu}}_\mu+\frac{1}{2}\left(({\xi^{\mu}}_\mu)^2-\xi^{\mu\nu}\xi_{\nu\mu}\right)-\frac{1}{6}\left[({\xi^{\mu}}_\mu)^3-3 {\xi^{\mu}}_\mu (\xi^{\rho\nu}\xi_{\nu\rho}))+2\left(\xi^{\mu\nu}\xi_{\nu\rho}{\xi^{\rho}}_{\mu}\right)\right]+ \det (\xi)~.
\end{align}
We notice that $\xi$ can be expressed as
\begin{align}
    \xi^{\mu\nu}= S^{\mu\nu}+ iA^{\mu\nu}~,
\end{align}
where $S$ is a totally symmetric part and $A$ a totally anti-symmetric part given by the expressions
\begin{align}
   & S^{\mu\nu}= \frac{1}{\zeta} \left(k^\mu k^\nu+ V^{\mu}V^\nu\right),\\
  &A^{\mu\nu}= \frac{1}{\zeta}(V^{\mu}k^\nu-V^\nu k^\mu)~,  
\end{align}
which greatly simplifies the upcoming computations. Term by term, the determinant will be
\begin{align}
 &{\xi^{\mu}}_\mu= \frac{k^2+V^2}{\zeta}~,\\
 &\xi^{\mu\nu}\xi_{\nu\mu}= \frac{(k^2+V^2)^2}{\zeta^2}~,\\
 &\xi^{\mu\nu}\xi_{\nu\rho}{\xi^{\rho}}_{\mu}= \frac{(k^2+V^2)^3}{\zeta^3}~, \\
 &\det\xi=0~,
\end{align}
this reduces the expression to
this reduces the expression to
\begin{align}
    \det( \eta-\xi)=1- {\xi^{\mu}}_\mu= 1- \frac{k^2+V^2}{\zeta}~,
\end{align}
yielding for $\det D$
\begin{align}
    \det D=& (-\zeta)^4\det(\eta-\xi)=\zeta^4 (1- \frac{k^2+V^2}{\zeta})\\
    =& \zeta^3(\zeta- k^2+V^2)= 0~,
\end{align}
showing that it is null, and by extension that the matrix is not invertible. This confirms our previous intuition about gauge invariance, and will require gauge fixings to obtain the propagator \cite{Itin:2009}. However, we can consider that our dispersion matrix possess the following Eigenvalue structure: 
\begin{align}
    D\{0,-\zeta,-\zeta,-\zeta\}
\end{align}
and thus our dispersion relation is simply characterised by $\zeta=0$. Doing so  we get
\begin{align}
    k^2+V^2=0
\end{align}
which will read 
\begin{align}
    \omega= \pm \sqrt{\|\vec k\|^2-V^\mu V_\mu} ,
\end{align}
indicating the presence of a rest frame, depending on the square of the background vector. Nevertheless, we can see that in the most general case, this is not necessarily defined to be real, notably if $V^\mu$ is defined to be purely time-like or when $\|\vec k\|^2 < {V^0}^2$. As such, the stability of the theory would impose 
\begin{align}\label{stability_cond}
    \|\vec k\|^2 \geq V^\mu V_\mu~.
\end{align}
Furthermore, requiring the Hamiltonian to be positive definite and the frequencies to be real and positive implies that $V^\mu$ must be space-like. This requirement is therefore consistent with the condition (\ref{stability_cond}). In the case where $V^\mu$ would be purely time-like, or that 
\begin{align}\label{instability_cond}
    \|\vec k\|^2 \leq V^\mu V_\mu~,
\end{align}
which would lead some infrared modes of the potential four vector to become unstable and grow exponentially. This process would, however, be fully unitary, and corresponds to a transfer of energy between the background field and the potential four vector.
We can underline the fact that, in our previous assumptions, this was already accepted to be the case, since we consider that $V^\mu$ is a logarithmic gradient of the background fields, and said fields are taken to be non-dynamical ({\textit{i.e.}} constant in time), ensuring the healthy behaviour of the model.  We can further highlight this by looking at the case where $ V^\mu$ is purely space-like and get
\begin{align}\label{dispersion_relation}
    \omega= \pm \sqrt{\|\vec k\|^2+\|\vec V\|^2}~.
\end{align}

Before moving on to the propagator, we shall derive the group velocities for the theory. To characterise more precisely how the massive mode propagates, we now introduce the phase and group refractive indices associated with the dispersion relation, following the standard characterization of wave
propagation in non-linear electrodynamics
\cite{Cruz:2025xut}. Starting from (\ref{dispersion_relation}) we get
\begin{align}
    \vec{v}_{g} =
    \frac{\partial \left|\omega\right|}{\partial \vec{k}} 
    =
    \frac{\vec{k}}{\sqrt{\|\vec k\|^2+\|\vec V\|^2}}~.
\end{align}

The corresponding refraction indices, in units $c = 1$
\begin{align}
    n_{g}
    =
    \frac{1}{\left\lVert \vec{v}_{g} \right\rVert}
    =
    \frac{\sqrt{\|\vec k\|^2+\|\vec V\|^2}}{\left\lVert \vec{k} \right\rVert}
    =
    \left(1+\frac{\|\vec V\|^2}{\|\vec k\|^2} \right)^{\frac{1}{2}}.
\end{align}

The phase velocity is defined by
\begin{align}
    v_{\phi}
    =
    \frac{\left|\omega\right|}{\left\lVert \vec{k} \right\rVert}
    =
    \frac{\sqrt{\|\vec k\|^2+\|\vec V\|^2}}{\left\lVert \vec{k}\right\rVert}
    =
    \left(1+\frac{\|\vec V\|^2}{\|\vec k\|^2} \right)^{\frac{1}{2}},
\end{align}
and the corresponding phase refractive index is thus
\begin{align}
    n_{\phi}
    =
    \frac{1}{v_{\phi}}
    =
    \left( 1+ \dfrac{\|\vec V\|^2}{\|\vec k\|^2}\right)^{-\frac{1}{2}}.
\end{align}

These expressions satisfy $n_\phi n_g = 1$, with $n_\phi < 1$ and $n_g > 1$. Accordingly, the phase velocity is superluminal, while the group velocity remains subluminal. This behaviour is consistent with the propagation of a massive dispersive mode and reflects the effective optical properties induced by the background field.

\subsection{Effective Mass scale and rest frequency}
Using the dispersion relation (\ref{dispersion_relation}), the module of the group velocity becomes in terms of $\omega$

\begin{align}
    \left\lVert \vec{v}_{g} \right\rVert 
    =
    \left( 1- \frac{\|\vec V\|^2}{\omega^2}\right)^{\frac{1}{2}}.
\end{align}

This new expression of the group velocity induces a frequency-dependent propagation velocity \cite{Goldhaber:1971mr}. It also implies the existence of a minimum frequency, that shall be called the \textit{cut-off frequency},  $\omega_{\mathrm{c}}=\lVert \vec{V} \rVert$. At this frequency, the group velocity is null and the wave is no longer propagating. This can be also interpreted as a mass gap introduced by the field $V$, since 
\begin{align}
    \omega^2
    =
    \|\vec k\|^2+\omega_{\mathrm{c}}^2 ~ 
    \Leftrightarrow 
    ~ E^2 = \vec{p}^2c^2 + m^2c^4,
\end{align}
taking $E= \hbar\omega$, and $\vec{p}= \hbar \vec{k}$, we get
\begin{align}
    \hbar^2 \omega^2
    = 
    \hbar^2 c^2\|\vec k\|^2 + m_{\mathrm{eff}}^2c^4,
\end{align}
that leads to
\begin{align}
    m_{\mathrm{eff}} = \frac{\hbar \omega_{\mathrm{c}}}{c^2}~.
\end{align}
In units $\hbar = c = 1$, it reads $m_{\mathrm{eff}} = \omega_{\mathrm{c}}$. Note that when $\omega = \omega_{\mathrm{c}}$, it is equivalent to $\vec{k} = \vec{0}$, which means that $\omega_{\mathrm{c}}$ can also be interpreted as the \textit{rest frequency}. In this mode, the rest energy is non-null
\begin{align}
    E(\vec{k} = \vec{0}) = \omega_{\mathrm{c}} = m_{\mathrm{eff}}.
\end{align}
Additionally, the wave number can be written as 
\begin{align}
    \lVert \vec{k} \rVert
    =
    \sqrt{\omega^{2}- \vec{V}^{\,2}} 
    = \sqrt{\omega^{2}- \omega_{\mathrm{c}}^2}~,
\end{align}
implying that propagating modes exist only for $\omega \geq \omega_{\mathrm{c}}$ since below this threshold, the wave number becomes
imaginary and the corresponding solutions are spatially evanescent
rather than propagating. This frequency dependence closely resembles that arising in de Broglie-Proca electrodynamics for a massive photon. Consequently, low-frequency electromagnetic waves provide enhanced sensitivity to the effective mass scale. In particular, radio observations of pulsars and fast radio bursts have been used to constrain a possible photon rest mass through frequency-dependent propagation delays \cite{wei-wu-2018, Bartlett:2021}. These bounds rely on the fact that, for a massive photon, lower-frequency electromagnetic waves propagate more slowly than higher-frequency ones over large astrophysical distances. The parameter $m_{\mathrm{eff}}$ appearing in our model could therefore, in principle, be constrained through analogous frequency-dependent propagation effects, in direct analogy with bounds obtained in de Broglie-Proca electrodynamics \cite{retino-spallicci-vaivads-2016, spallicci-etal-2024b}.

\section{Modified transverse operator and propagator}
Going back to the dispersion matrix \eqref{dispersionmatrix}
\begin{align}
    D^{\mu\nu}=-\zeta \left(\eta^{\mu\nu}- \frac{1}{\zeta}\chi^{\mu\nu}\right)~,
\end{align}

 we can consider that it defines a modified transverse operator, which we will rename $\mathcal{P}^{\mu\nu}$
\begin{align}
    D^{\mu\nu}=-(k^2+V^2)\tilde{\mathcal{P}}^{\mu\nu}(k) = (\cev{\partial}_\alpha -
V_\alpha)(\vec{\partial}^\alpha -V^\alpha) \mathcal{P}^{\mu\nu}(x)
\end{align}
with 
\begin{align}
    &\tilde{\mathcal{P}}^{\mu\nu}(k)= \left[\eta^{\mu\nu}- \frac{(k^\mu+iV^\mu)(k^\nu-iV^\nu)}{k^2+V^2}\right]~,\\
    &\mathcal{P}^{\mu\nu}(x)= \left[\eta^{\mu\nu}+ \frac{(\partial^\mu+V^\mu)(\partial^\nu-V^\nu)}{(\cev{\partial}_\alpha -
V_\alpha)(\vec{\partial}^\alpha -V^\alpha)}\right]~,
\end{align}
 which renders the dispersion matrix symmetric when derivatives act symmetrically to the left and right, and the single-derivative terms may be rewritten as
\begin{align}
    \partial_\mu 
    =
    \frac{1}{2} (\cev{\partial}_\mu - \vec{\partial}_\mu)
\end{align}
so they naturally appear antisymmetric.

We can however quickly see that said operator does not admit $k ^\mu$ as a null vector, further confirming that the wave vector by itself is not transversal to the propagation. We can however find naïvely the null vector of the model, by comparing it with Maxwell's transversal projector
\begin{align}
    P^{\mu\nu}=\eta^{\mu\nu}-\frac{k^\mu k^\nu}{k^2}~,
\end{align}
and noticing that the $k^\mu$ vectors, are replaced by their shifted counterparts $k^\mu\pm iV^\mu$. The imaginary factor here, is not indicating a complexified gauge direction, but is simply an artifact of the phase exponential ansatz used earlier on. Doing this replacement we can see that
\begin{align}
    \mathcal{P}^{\mu\nu}(k_\nu+iV_\nu)=0~,
\end{align}
which indicates that the null vector of said theory is $\mathcal{K}^\mu= k^\mu+i V^\mu$. Furthermore, we can see that 
\begin{align}
    \mathcal{K}^2=(k^\mu+iV^\mu)(k_\mu-iV_\mu)= k^2+ V^2=0~,
\end{align}
which furthers confirms the genuine modification of the gauge structure of the theory, as well as the transversal nature of the mass term that was hinted in \cite{Dib:2025qoy} through the propagator.
\subsection{Derivation of the propagator}
To obtain said propagator, we will have to invert the ${D^{\mu\nu}}$ operator. Starting from \eqref{action 1}
\begin{align}
    \mathcal{L}= -\frac14f'^{\mu\nu}f'_{\mu\nu}-a'_\mu V_\nu f'^{\mu\nu}-\frac12 \left(a'^\nu a'_\nu V^\mu V_\mu- a'_\mu a'_\nu V^\mu V^\nu\right)~~,
\end{align}
which we rewrite as
\begin{align}
    \mathcal{L}= -\frac{1}{4} f'^2+ V_\mu a'_\nu \partial ^\mu a'^\nu-V_\mu a'_\nu \partial^\nu a'^\mu - \frac{1}{2}V^2 a'^2+ \frac{1}{2}V_\mu V_\nu a'^\mu a'^\nu~.
\end{align}
 The second term can be expressed as a surface term, and we may rewrite the third one as
\begin{align}
    -V_\mu a'_\nu\partial^\nu a'^\mu=& -\frac12 V_\mu a'_ \nu \partial^\nu a'^\mu-\frac12 V_\mu a'_ \nu \partial^\nu a'^\mu \nonumber\\
    =& \frac12 a'^\mu \partial _\nu (V_\mu a'^\nu)- \frac12 V_\nu a'_\mu \partial ^\mu a'^\nu \nonumber\\
    &=\frac12 a'^\mu(V_\mu \partial_\nu- V_\nu \partial_\mu) a'^\mu~. 
\end{align}
By then decomposing the metric as 
\begin{align}
    \eta^{\mu\nu}= \theta^{\mu\nu}+\omega^{\mu\nu}~,
\end{align}
where $\theta^{\mu\nu}$ is the standard transverse projector and $\displaystyle\omega^{\mu\nu}=\frac{\partial^\mu \partial^\nu}{\Box}$ the longitudinal projector, the first term becomes
\begin{align}
    -\frac14 f'^2=\frac12 a^\mu \Box \theta_{\mu\nu} a^\nu~,
\end{align}
since the field strength is a transverse function \cite{itzykson-zuber-2006, Weinberg:1995mt}. After some manipulations the action will read
\begin{align}
   &\mathcal{L}= \frac{1}{2}a^\mu \left[ \left(\Box- V^2\right)\theta_{\mu\nu} - V^2 \omega_{\mu\nu}+V_\mu\partial_\nu -V_\nu \partial_\mu + V_\mu V_\nu\right]a^\nu ~,\\
   &\mathcal{L}= \frac{1}{2}a^\mu M_{\mu\nu}a^\nu~,
\end{align}
where $M^{\mu\nu}$ is given by
\begin{align}
    M_{\mu\nu}=\alpha \theta_{\mu\nu}+\beta \omega_{\mu\nu}+ \gamma V_\mu \partial_\nu+ \delta V_\nu \partial_\mu+ \epsilon V_\mu V_\nu~,
\end{align}
with
\begin{align}
    &\alpha= \Box- V^2\\
    &\beta= -V^2\\
    &\gamma=-\delta=1\\
    &\epsilon=1~.
\end{align}

Our task is now to find $M^{-1}$ such as
\begin{align}
    M^{\mu\kappa}M^{-1}_{\kappa\nu}= \delta^{\mu}_\nu={\theta^{\mu}}_\nu + {\omega^\mu }_\nu~.
\end{align}
We start by developing the above expression
\begin{align}
   &\left(\alpha \theta^{\mu\kappa}+\beta \omega^{\mu\kappa}+ \gamma V ^\mu \partial^\kappa+ \delta V^\kappa \partial^\mu+ \epsilon V^\mu V^\kappa\right)\left(x \theta_{\kappa\nu}+y \omega_{\kappa\nu}+ z V_\kappa \partial_\nu+w V_\nu \partial_\kappa+ t V_\kappa V_\nu\right)= \delta^{\mu}_{\nu} ~,
\end{align}
and by expanding every term on the left hand side
\begin{align}
    &\alpha x ~{\theta^{\mu}}_{\nu}+ \alpha z \left(V^\mu \partial_\nu -V^{\kappa}\partial_\kappa {\omega^{\mu}}_\nu \right)+ \alpha t \left(V^\mu V_\nu-\frac{\partial^\kappa V_\kappa}{\Box}\partial^\mu V_\nu\right)+ \beta y {\omega^{\mu}}_\nu+ \beta z V^\kappa \partial_\kappa {\omega^{\mu}}_\nu+ \beta w \partial^\mu V_\nu+\nonumber\\
    &\beta t \frac{V^k\partial_k}{\Box}\partial^\mu V_\nu \gamma y  V^\mu \partial_\nu+\gamma z V^\kappa \partial_\kappa V^\mu \partial_\nu+\gamma w \Box V^\mu V_\nu+\gamma t V^\kappa\partial_ \kappa V^\mu V_\nu+\delta x \left(\partial^\mu V_\nu-V^{\kappa}\partial_\kappa {\omega^{\mu}}_\nu\right)+\nonumber\\
 &\delta y V^{\kappa}\partial_\kappa{\omega^{\mu}}_\nu+ \delta z V^2 \Box{\omega^{\mu}}_\nu\delta w V^{\kappa}\partial_\kappa \partial^\mu V_\nu+ \delta t V^2\partial^\mu V_\nu \epsilon x \left(V^\mu V_\nu-\frac{V^\kappa \partial_\kappa}{\Box}V^\mu \partial_\nu\right)+\epsilon y \frac{V^{\kappa}\partial_\kappa}{\Box}V^{\mu}\partial_\nu \partial_\nu \nonumber\\
&+ \epsilon z V^2 V^\mu+\epsilon w V^{\kappa}\partial_\kappa V^{\mu}V_\nu+\epsilon t V^2V^\mu V_\nu={\theta^{\mu}}_\nu +{\omega^{\mu}}_\nu~,
\end{align}
we get the following set of equations
\begin{align}
    &{\theta^\mu}_\nu \rightarrow \alpha x=1 \rightarrow x=\frac{1}{\alpha}~,\\
    &{\omega^{\mu}}_\nu \rightarrow -\alpha z V^{\kappa}\partial_\kappa+ \beta y+ \beta z V^{\kappa}\partial_\kappa-\delta xV^{\kappa}\partial_\kappa+ \delta y V^{\kappa}\partial_\kappa+\delta z V^2 \Box=1~, \\
    &V^\mu \partial_\nu \rightarrow \alpha z+ \gamma y+ \gamma z V^{\kappa}\partial_\kappa- \epsilon x \frac{V^{\kappa}\partial_\kappa}{\Box}+ \epsilon y\frac{V^{\kappa}\partial_\kappa}{\Box}+\epsilon z V^2=0~,\\
    &V_\nu \partial^\mu \rightarrow-\alpha t \frac{V^{\kappa}\partial_\kappa}{\Box}+\beta w + \beta t\frac{V^{\kappa}\partial_\kappa}{\Box}+\delta x + \delta w V^{\kappa}\partial_\kappa+\delta tV^2=0~,\\
    &V^{\mu}V_\nu \rightarrow\alpha t +\gamma w \Box + \gamma t V^{\kappa}\partial_\kappa+\epsilon x + \epsilon w V^{\kappa}\partial_\kappa+ \epsilon t V^2=0~.
\end{align}

We can further decompose them into two independent systems of linear equations
\begin{align}
    &(\beta+\delta V^{\kappa}\partial_\kappa)y+(-\alpha V^{\kappa}\partial_\kappa+\beta V^{\kappa}\partial_\kappa+\delta V^2 \Box)z=1+\frac{\delta V^{\kappa}\partial_\kappa}{\alpha}\nonumber\\
    &(\gamma +\epsilon \frac{V^{\kappa}\partial_\kappa}{\Box})y +(\alpha+\gamma V^{\kappa}\partial_\kappa+ \epsilon V^2)z=\frac{\epsilon V^{\kappa}\partial_\kappa}{\alpha \Box}~,
\end{align}
and 
\begin{align}
    &(\beta + \delta V^{\kappa}\partial_\kappa) w+ (-\alpha \frac{V^{\kappa}\partial_\kappa}{\Box}+\beta \frac{V^{\kappa}\partial_\kappa}{\Box}+\delta V^2)t= -\frac{\delta}{\alpha}\nonumber\\
    &(\gamma \Box + \epsilon V^{\kappa}\partial_\kappa) w +(\alpha+\gamma V^{\kappa}\partial_\kappa+ \epsilon V^2)t =-\frac{\epsilon}{\alpha}~.
\end{align}
We must then re-express them as matrices and find their determinant to move forward. We will skip the demonstration, and write the resulting determinant as
\begin{align}
    \det = \alpha\left(\beta+ (\gamma+\delta)V^{\kappa}\partial_\kappa+\epsilon\frac{(V^{\kappa}\partial_\kappa)^2}{\Box}\right)+\left(\frac{\beta \epsilon}{\Box}-\gamma \delta\right)\left(V^2 \Box-(V^{\kappa}\partial_\kappa)^2\right) \equiv \Delta~,
\end{align}
and finally getting
\begin{align} \label{propagator}
    {M}^{-1}_{\mu\nu}= \frac{1}{\alpha}\theta_{\mu\nu}&+\frac{1}{\Delta}\left[\left(\alpha+\gamma V\cdot\partial+\epsilon V^2\right)\left(1+\frac{\delta}{\alpha}V\cdot\partial\right)+\left((\alpha-\beta)V\cdot\partial-\delta V^2\Box\right)\frac{\epsilon}{\alpha}\frac{V\cdot \partial}{\Box}\right]\omega_{\mu\nu}&\nonumber\\
    &+\frac{V_\mu \partial_\nu}{\Delta}\left[\left(-\gamma-\epsilon\frac{V\cdot\partial}{\Box}\right)\left(1+\frac{\delta}{\alpha}V\cdot\partial\right)+\left(\beta+\delta V\cdot \partial\right)\frac{\epsilon}{\alpha} \frac{V\cdot \partial}{\Box}\right]\nonumber\\
    &+\frac{V_\nu \partial_\mu}{\Delta}\left[-\left(\alpha+\gamma V\cdot \partial+\epsilon V^2\right)\frac{\delta}{\alpha}-\left(\alpha \frac{V\cdot\partial}{\Box}-\beta \frac{V\cdot\partial}{\Box}-\delta V^2 \right)\frac{\epsilon}{\alpha}\right] &\nonumber\\
    &+\frac{V_\mu V_\nu}{\Delta}\left[\left(\gamma \Box+ \epsilon V\cdot\partial\right)\frac{\delta}{\alpha}-\left(\beta+ \delta V\cdot\partial\right)\frac{\epsilon}{\alpha}\right].
\end{align}

By now recalling the definition of our coefficients, we can see two distinct poles for the propagator: One on the transverse part
\begin{align}
    \frac{1}{\alpha}= \frac{1}{\Box- V^2} \rightarrow \frac{1}{k^2+V^2}~,
\end{align}
which as stated in \cite{Dib:2025qoy} confirms the transversal nature of the mass. The second pole is obtained from the longitudinal part 
\begin{align}
    \frac{1}{\Delta}~,
\end{align}
with $\Delta$
\begin{align}
    \Delta= (\Box - V^2) \left[- V^2 +\frac{(V\cdot\partial)^2}{\Box}\right]+ (-\frac{V^2}{\Box}+1)(V^2\Box- (V\cdot \partial)^2)=0~,
\end{align}
showing that the longitudinal part is singular. This is a manifestation of the gauge invariance of the theory, and we must introduce a gauge fixing to solve this. In the action this reads
\begin{align}
    \mathcal{L}_{gf}=\frac{1}{2 \xi} \left(\partial_\mu a'^\mu\right)^2 =-\frac{1}{\xi}  a'^\mu\Box \omega_{\mu\nu} a'^\nu~,
\end{align}
this in turn changes the $\beta$ coefficient such as to have
\begin{align}
    \alpha=\Box-V^2 ,~~\beta= -\frac{\Box}{\xi}-V^2,~~\gamma=-\delta=\epsilon=1~,
\end{align}
which finally gives us for $\Delta$
\begin{align}
    \Delta= -\frac{1}{\xi}\left(\Box^2+\left(V\cdot\partial\right)^2\right)~.
\end{align}
We may now confidently state that there exist two poles 
\begin{itemize}
    \item One on the transverse part $\theta_{\mu\nu}$ by setting $\alpha=0$ which gives
    \begin{align}
        \alpha=0 \Rightarrow k^2=-V^2= \|\vec V\|^2~.
    \end{align}
    \item And a second one on the longitudinal part $\omega_{\mu\nu}$ by solving $\Delta=0$ which reads
    \begin{align}
        k^4-\left(V\cdot k\right)^2=(k^2+V\cdot k)(k^2-V\cdot k)= \begin{cases}
            k^2+ \vec{V}\cdot\vec{k}=0~,\\
            k^2- \vec{V}\cdot\vec{k}=0
        \end{cases}.
    \end{align}
\end{itemize}
The presence of the massive pole on the transverse part is reminiscent of topologically induced masses \cite{chernsimons1974, Dvali-Jackiw-Pi-2006,Deser-Jackiw-Templeton-1982}, meaning that the mass generated by the background field has to be accommodated by the null helicity component of the spin-1 carried by the photon. 

\section{Canonical analysis and gauge structure}
\subsection{Hamiltonian picture }
To rule definitely on the gauge nature of the theory, we must proceed with its Hamiltonian analysis and classification of its constraints \cite{Henneaux-Teitelboum-1992}. Starting from the action
\begin{align}
    \mathcal{L}= -\frac{1}{4} f'^2+ V_\mu a'_\nu \partial ^\mu a'^\nu-V_\mu a'_\nu \partial^\nu a'^\mu - \frac{1}{2}V^2 a'^2+ \frac{1}{2}V_\mu V_\nu a'^\mu a'^\nu~,
\end{align}
and breaking its manifest covariance 
\begin{align}
    \mathcal{L}=-\frac{1}{2} {f'_{0i}}^2 -\frac{1}{4}{f'_{ij}}^2-(a'_0V_i-a'_i V_0) f'^{0i}-a'_i V_j f'^{ij}- \frac{1}{2} \left({a'_0}^2 V_i ^2+ {a'_i}^2 V^\mu V_\mu-2a'_ia'_0 V^0 V^i-a'_i a'_j V^i V^j\right)~,
\end{align}
before performing the Legendre transform to obtain the conjugate momenta
 \begin{align}
  & \pi'^\mu =\frac{\partial \mathcal{L_0}}{\partial \dot{a'_\mu}}=\! \begin{cases}
       & \!\!\!\!\! \pi'^0= \frac{\ds\partial \mathcal{L} }{\ds\partial \partial_0 a'_0}=0~,\\
      & \!\!\!\!\! \pi'^i= \frac{\ds\partial \mathcal{L}}{\ds\partial \partial_0 a'_i}=f'^{i0}-(a'^0V^i-a'^i V^0) ~,\nonumber
    \end{cases}&
\end{align}
where $\pi'^0=0$ is the primary constraint, and will be relabelled as $\phi_1= \pi'^0=0$.

The canonical Hamiltonian will then read
\begin{align}
    \mathcal{H}_C= \int d^3 \vec{x} \left[-a'_0\partial^i \pi'_i - \frac12 \pi'^i\pi'_i -\pi'^i\left( a'_0V_i-a'_iV_0\right)+\frac14 f'_{ij} f'^{ij}+ a'_i V_j f'^{ij}-\frac12 \left(\left(a'_i V^i\right)^2- \left(a'_i V_j\right)^2 \right)\right]~.
\end{align}
However, since we are dealing with constrained Hamiltonian systems, the time evolution will be dictated by the total Hamiltonian, which is defined as
\begin{align}
    \mathcal{H}_T= \mathcal{H}_C+ \sum_{i=1}^{i=N}\lambda_i \phi^i~,
\end{align}
where $\lambda_i$ are Lagrange multipliers that enforce the primary constraints of the theory. In our case, this gives
\begin{align}
    \mathcal{H}_T= \int d^3 \vec{x} \left[-a'_0\partial^i \pi'_i - \frac12 \pi'^i\pi'_i -\pi'^i\left( a'_0V_i-a'_iV_0\right)+\frac14 f'_{ij} f'^{ij}+ a'_i V_j f'^{ij}-\frac12 \left(\left(a'_i V^i\right)^2- \left(a'_i V_j\right)^2 \right)+ \lambda_1 \phi_1\right]~.
\end{align}

 We must then ensure the consistency of our constraints, and to do so we will proceed with the Dirac Bergmann algorithm \cite{Anderson-Bergmann-1951, Bergmann-Goldberg-1955,Dirac-1950} as follows :\begin{itemize}
    \item 1) Evaluate the Poisson Bracket between the primary constraint and the total Hamiltonian.
    \item 2) If this does not fixes a Lagrange multiplier, or yields a trivial equality, it has generated another constraint, then repeat step one with this additional constraint. 
    \item 3) Stop once there no new additional constraints.
\end{itemize}

For our primary constraint $\phi_1=0$, this gives
\begin{align}
    \{\phi_1, \mathcal{H}_T\}= \partial^i\pi'_i + \pi'^i V_i \approx0~,
\end{align}
which will be the modified Gauss law for theory, relabelled as $\phi_2= \partial_i \pi'^i+ \pi'^i V_i \approx0$. The Poisson Bracket between the Gauss law and the total Hamiltonian will read
\begin{align}
    \{\phi_2,\mathcal{H}_T\}=\{\partial_i \pi'^i, \pi'^i a'_i V_0\}+\{\pi'^iV_i, \pi'^i a'_ i V_0\}~,
\end{align}
where all of the other brackets cancel each other, leaving us with this combination as the only non trivial contribution. This reduces to
\begin{align}
    \{\phi_2, \mathcal{H}_T\}=& V_0 \partial_i \pi'^i+ \pi'^i V_i V_0 \approx0\nonumber\\
    =& V_0 \phi_2~,
\end{align}
showing that this does not generate an additional constraint as it can be expressed as a previously derived constraint. The Dirac Bergmann algorithm stops, and we may state definitely that the theory is composed of two constraints $\phi_1, \phi_2$.

The classification of the constraints is trivial and will immediately give
\begin{align}
    \{\phi_1,\phi_2\}=0~,
\end{align}
showing the presence of two first class constraints, confirming the gauge theoretical aspect of the theory. Additionally, this also confirms our earlier hunch, that fixing the totally symmetric and traceless part of the $V^{\mu} V^\nu$ tensor in \cite{Dib:2025qoy} acted effectively as an ill-suited gauge fixing.

The physical degrees of freedom of the theory are given by the master formula \cite{Henneaux-Teitelboum-1992} 
\begin{align}
& 2\!\times\! DOF = ~ \text{Phase Space variables}\!-\! 2 \!\times\! 1^{\text{st}}\text{Class Constraints}- 2^{\text{nd}}\text{Class Constraints}& \label{MasterFormulaDOF}\\
& 2\!\times\! DOF= 8-2\times 2 -0~~~ \Rightarrow DOF= 2\nonumber~,
\end{align}
which highlights the absence of additional polarisations. 

The extended Hamiltonian, which includes all first class constraints, is however still dependent on two undefined Lagrange multipliers
\begin{align}
    \mathcal{H}_{ext}= \mathcal{H}_C+\int d^3 \vec{x}~ \lambda_1\phi_1+ \lambda_2 \phi_2~.
\end{align}
Let us then consider the evolution of the time-like polarisation 
\begin{align}
    \{a'_0, \mathcal{H}_T\}= \lambda_1~,
\end{align}
which confirms the arbitrariness of the scalar potential, allowing us to set it to zero. 

Since we have set $a'_0=0$, we may inject this result in the total Hamiltonian and check its positivity. We start with
\begin{align}
    \mathcal{H}_T=\int d^3\vec{x}\left[ -\frac12 \pi'^i\pi'_i+\pi'^i a'_iV_0+ \frac14 f'_{ij}f'^{ij}+a'_i V_j f'^{ij}-\frac12 \left\{ \left(a'_iV^i\right)^2-\left(a'_iV_ j\right)^2\right\}+\lambda_1 \pi'^0\right]~,
\end{align}
and then proceed to anti-symmetrise the fourth term as 
\begin{align}
    a'_i V'_j f'^{ij}= \frac12(a'_iV_j-a'_jV_i)f'^{ij}=\frac12 V_{ij}f'^{ij}~,
\end{align}
where $V_{ij}=a'_iV_j-a'_jV_i$. With this new 2-form in hand, we notice that the second to last term in the Hamiltonian can be expressed exactly as
\begin{align}
    \frac14V_{ij}V^{ij}=\frac12\left(\left[a'_i V_j\right]^2-\left[a'_iV^i\right]^2\right)~,
\end{align}
recasting the total Hamiltonian as
\begin{align}
    \mathcal{H}_T=\int d^3\vec{x}\left[ -\frac12 \pi'^i\pi'_i+\pi'^i a'_iV_0+ \frac14 f'_{ij}f'^{ij}+\frac12V_{ij} f'^{ij}+\frac14 V_{ij}V^{ij}+\lambda_1 \pi'^0\right]~.
\end{align}
Focusing now on the canonical Hamiltonian, we can show that its expression will then read
\begin{align}
    \mathcal{H}_C=\int d^3\vec{x}\left[ \frac12 \left(\pi'^i\right )^2+\pi'^i a'_iV_0+ \frac14 \left(f'^{ij}+V^{ij}\right)^2\right]~,
\end{align}
which for purely space-like $V^\mu$ reduces to the following expression
\begin{align}
     \mathcal{H}_C=\int d^3\vec{x}\left[ \frac12 \left(\pi'^i\right )^2+ \frac14 \left(f'^{ij}+V^{ij}\right)^2\right]~,
\end{align}
that is positive definite for arbitrary background fields. We notice that in the purely space-like case, the background field causes a shift in the magnetic part of the action.
\subsection{Gauge transformation and structure}
The vector potential $a'^i$ is not fully defined through its evolution due to the remaining multiplier
\begin{align}
   \delta a'^i= \{a'^i, \int d^3\vec{x} ~\lambda_2 (\partial_j \pi^j+ \pi^j V_j)\}= -(\partial^i-V^i) \lambda_2~,
\end{align}
which is the expression of our local gauge transformation. This shows that the background vector field shifts the derivative acting on the fields by a constant value, letting us express the form of our covariant derivative as
\begin{align}
    D_\mu= \partial_\mu- V_\mu~,
\end{align}
which can be further confirmed by verifying Noether's second theorem. Indeed, if the covariant divergence of the equation of motion vanishes identically then the corresponding covariant derivative defines a Noether identity \cite{Avery:2015rga} and therefore generates a local gauge symmetry. Considering \eqref{EOM'1}
\begin{align}
    \mathcal{E}^\nu=\partial_\mu f'^{\mu\nu}+ V^\nu \partial^\mu a'_\mu - V^\mu \partial^\nu a'_\mu-a'^{\nu} V^\mu V_\mu+ a'^\mu V^{\nu}V_\mu~~,
\end{align}
we can see that at the level of the action
\begin{align}
    S=\int d^4 x~ \mathcal{E}^\mu a_\mu~,
\end{align}
and after transformation of $a^\mu$ as
\begin{align}
    \delta a^\mu= (\partial_\mu-V_\mu) \Lambda~,
\end{align}
it then becomes
\begin{align}
    \delta S= \int d^4x ~\mathcal{E}^\mu \left(\partial_\mu-V_\mu\right) \Lambda~.
\end{align}
After integration by parts and discarding the surface term we get
\begin{align}
    \delta S= -\int d^4 x \Lambda \left(\partial_\mu+V_\mu\right) \mathcal{E}^\mu~,
\end{align}
therefore
\begin{align}
    \bar{D}_\nu \mathcal{E}^{\nu}=(\partial_\nu+V_\nu) \mathcal{E}^\nu=0~.
\end{align}

The fact that the local gauge transformation is of the form discussed above ({\textit{i.e.}} the usual U(1) transformation + constant parameter) is reassuring, as it implies that the algebra formed by the first class constraints still closes, ensuring the nilpotency of the BRST operator \cite{Henneaux-Teitelboum-1992}. In fact, one could go further, and show that said algebra is still Abelian. Indeed, considering 
\begin{align}
    \delta_\Lambda a^\mu= D^\mu \Lambda~,
\end{align}
we can compute 
\begin{align}
    \left[\delta_{\Lambda_1}, \delta_{\Lambda_2}\right]a^\mu=\delta_{\Lambda_1}\delta_{\Lambda_2}a^\mu-\delta_{\Lambda_2}\delta_{\Lambda_1}a^\mu~,
\end{align}
keeping in mind that the gauge parameters are independent $\delta_{\Lambda_2}\Lambda_1=0$, it is trivial to show that the gauge algebra is Abelian
\begin{align}
    \left[\delta_{\Lambda_1}, \delta_{\Lambda_2}\right]=0~,
\end{align}
exactly as Maxwell's case, thus not influencing the BRST structure of the model.

 We may now rule definitely on the gauge structure of the theory and confirm what was speculated in \cite{Dib:2025qoy}, which is that the field redefinition described in \eqref{fieldredif} is only a formal symmetry breaking, and only the manifest gauge invariance has been lost, in the same manner that a Higgs mechanism would act on the vacuum state. \\

In section (2.2), we made some simplifying assumptions that were twofold: first, we decided to ignore higher order contributions to the action, and second we decided to drop dual terms. With the mass generation effect we uncovered, we might be worried that these terms would lead to modulation, or worst, cancel the mass generated. We will try to give some arguments to reassure the reader that even in the worst case scenario, the mass can only be modified, but never erased by these contributions. Let us begin with $D$ terms, and use for illustration $D_1$. We remind ourselves that it is defined as
\begin{align}
   D_1= \frac{\partial^2\mathcal{L}}{\partial \mathcal{F}^2}=\frac{\partial}{\partial \mathcal{F}}C_1=\frac{\partial}{\partial \mathcal{F}} \alpha^2~.
\end{align}
This term will be effectively incorporated in the action coupled to
\begin{align}
    \delta \mathcal{L}&=\frac18D_1 F^{\mu\nu}F^{\rho\sigma}f_{\mu\nu}f_{\rho\sigma}~\nonumber\\
    &= \partial_\mathcal{F} \alpha^2 F^{\mu\nu}F^{\rho\sigma}f_{\mu\nu}f_{\rho\sigma}~,
\end{align}
meaning that if there is any contribution to the mass, it should be coming from the field redefinition of $\partial_{\mathcal{F}} ~\alpha^2$ into the potentials, which will yield objects of dimensionality different to that of $V^\mu$. This implies that while this contribution might shift or correct the massive pole, it cannot cancel it identically.

As for the $C_2$ terms, let us suppose that they remained, and that the first one would be written as
\begin{align}
   -\frac14 C_2 \tilde{f^{\mu\nu}}f_{\mu\nu}=-\frac14 \beta^2 \tilde{f}f=-\frac14 \beta^2 \varepsilon^{\mu\nu\rho\sigma}f_{\mu\nu}f_{\rho\sigma}~.
\end{align}

If we proceed with the same field redefinition as a above, we will get 
\begin{align}
    \beta^2 \varepsilon^{\mu\nu\rho\sigma}f_{\mu\nu}f_{\rho\sigma}=\varepsilon^{\mu\nu\rho\sigma}f'_{\mu\nu}f'_{\rho\sigma}+ 4\varepsilon^{\mu\nu\rho\sigma} a'_\mu B_\nu f'_{\rho\sigma}+2 \varepsilon^{\mu\nu\rho\sigma}\left(a'_\mu a'_\rho B_\nu B_\sigma- a'_\mu a'_\nu B_\rho B_\sigma\right)~,
\end{align}
where $\displaystyle B^\nu= \frac{\partial^\nu \beta}{\beta}$ similarly to $ V^\mu$. We immediately see that the last term is identically zero, while the first one will be topological term. The only remaining component is the second one, which is a Carroll-Field-Jackiw term, which may modulate the mass and dispersion relations, but cannot destroy the massive behaviour since our mass term is quadratic in the action.

\section{Discussion}
It is usually accepted in physics that symmetry breaking (soft or spontaneous) leads to the generation of mass in gauge theories. Indeed, the breaking of one of the symmetries (gauge symmetry, chiral symmetry etc.) of a model, will turn generate Goldstone bosons which can be then "swallowed" by the particles to add a third polarisation to the theory, avoiding a discontinuity of degrees of freedom, and unitarity questions that would arise for non-Abelian theories \cite{Weinberg:1995mt,Peskin:1995ev,itzykson-zuber-2006}.

We can, however, have effective mechanisms such as topological mass generations (as discussed for example in the Chern-Simons \cite{chernsimons1974,Dvali-Jackiw-Pi-2006,Deser-Jackiw-Templeton-1982} theory in 2+1 dimensions) which will introduce a mass gap in the model without the addition of any degree of freedoms while preserving gauge invariance. In our case, the field redefinition at play appears to be generating a mass, and it also appears that, in analogy with Chern-Simons electrodynamics, said mass did not generate an additional degree of freedom. One key difference here (besides the obvious fact that Maxwell-Chern-Simons is a topological field theory, and ours is not), is that while our model preserves a form of gauge invariance, it induces through our assumptions a form of Lorentz symmetry violations, which tampers with Wigner's classification, making the analogy with a de Broglie-Proca type mass ill-suited, as such our theory should not be expected to follow the same construction as any type of Maxwell-Proca model \cite{Diez-etal-2020}. In a follow-up study, we shall consider the Little Group associated to the theory, and derived the deformed Casimir operators, to state more formally on the differences between what we are studying, and a standard massive particle. 

Let us nevertheless consider the following points: the coefficients coupling to the photon fields are not constants, which means that these dielectric coefficients are functions of space-time (in our case, space more specifically). This in turn has some consequences on the vacuum state of the system, which will develop a different refractive index and break its own symmetry. As such, while the original action seemed to obey two $U(1)$ invariances (photon and background), it is reasonable to believe that due to the non-dynamical nature of the background, a spontaneous symmetry breaking hidden at the quantum level would emerge. By choosing to absorb the order zero background contribution into the potential 4-vector, we are indirectly showing this effect by proceeding to a manifest symmetry breaking, which allows us to describe classically the otherwise subtle effects of the background field.  These mechanisms seem very close to those discussed in \cite{baetaetal2004} which is coherent since the constant nature of $V^\mu$ indicates the presence of a Lorentz symmetry violation in our model as stated above. This mechanism is however \textbf{not} a symmetry breaking sense since gauge invariance is preserved at every stages of the work, but the supposed modification of the rotation group leads to
the generation of massive poles without additional degrees of freedom.  

 Another analogy can be made in cosmology with the emergence of mass in the case of a massless scalar field in an expanding FLRW background \cite{Mukhanov:2005sc,Janssen:2009nz}. Applying a canonical field rescaling similar to (\ref{fieldredif}) converts the friction-like term into an effective mass-like term without breaking the underlying symmetry. For instance, a massless minimally coupled scalar field in a spatially flat FLRW spacetime, written in conformal time, satisfies

\begin{align}
\left(\partial_\eta^2 + 2\mathcal{H}\partial_\eta -\nabla^2\right)\phi 
=
0,
\qquad
\mathcal{H} \equiv \frac{a'}{a},
\end{align}
with $\mathcal{H}$ is the conformal expansion rate. Introducing the canonically normalized variable $(\phi \rightarrow u = a\phi)$ removes the first-derivative term and gives
\begin{align}
    \left(\partial_\eta^2 - \nabla^2 - \mathcal{H}^2 - \mathcal{H}'\right) u 
    =
    0.
\end{align}
This standard canonical rescaling in an expanding universe removes the friction of the physical field $\phi$ and replaces it by a mass term for the rescaled field $u$ 
\begin{align}
    m_{\rm eff}^2
    =
    -(\mathcal{H}^2+\mathcal{H}').
\end{align}
For spacetimes with constant deceleration parameter $\epsilon=-\dot H/H^2$ one has 
\begin{align}
    \mathcal{H}' = (1-\epsilon)\mathcal{H}^2, 
    \qquad
    m_{\rm eff}^2 = -(2-\epsilon)\mathcal{H}^2.
\end{align}

This provides an explicit analogy for the present Generalised electrodynamics: in the original variables $(A_\mu,a_\nu)$, the gauge structure is manifest and the background produces a friction-like modification of $(a_\mu)$ dynamics, whereas after the rescaling $(a_\mu \rightarrow a'_\mu)$, the friction is traded for a mass-like term. The latter should be interpreted with care, since its appearance is tied to the choice of canonical variables rather than, by itself, demonstrating a genuine breaking of the original gauge symmetry, which would induce a third polarisation for the theory. 

The clearest result that we can take away from this construction is that even though gauge symmetry is preserved, and the number of physical degrees of freedom is preserved, there is generation of a massive pole in the dynamical sector of the photon field, which is the strictest sense of a mass generation. Since the theory predicts a pole shift, this will have a consequence for the photon propagation, in particular the dynamical photon will propagate as a massive particle, which will limit the range of the electromagnetic interaction to $\sim m_\gamma^{-1}\sim 1/\sqrt{\|\vec V\|^2}$.

The unanswered question here is considering the physicality of the mass. Indeed, we showed that classically, the theory is well behaved (from the dispersion relations, poles, and Hamiltonian in pure space-like case), but quantisation can introduce anomalies \cite{Henneaux-Teitelboum-1992, Weinberg:1995mt} and might cast shadow on the massiveness of the model. The next step is to couple fermions to the theory, which might not be as trivial as it seems. Indeed, finding a fermionic representation might appear simple, 
but we have to keep in mind that if the effects of the background on photons are not trivial, they should also be non trivial on fermions. Once fully quantised, and with the proper Feynman rules in hand, we could finally state if our mass is physical, or simply an effective artifact that only holds at the classical level.

\section{Conclusions} 
In this paper, we studied in details the propagation features and canonical structure for the model introduced initially in \cite{Dib:2025qoy}. We showed that contrary to what was initially assumed, the theory does not exhibit a third polarisation, but still possesses a mass gap as highlighted from the pole structures and dispersion relations. From said dispersion relations and poles, we show that the effective mass is healthy, and the theory stable under reasonable conditions (such as selecting only space-like backgrounds). We showed within the canonical picture that the constraint algebra of the theory is composed of two first class constraints, generating a gauge transformation, that is akin to a deformed $U(1)$. We showed that said transformations still form an Abelian group, ensuring the nilpotency of the BRST operator associated to them. The Hamiltonian of the theory is also shown to be positive definite in the case of purely space-like background fields, with a modification of the magnetic sector. 

\section{Acknowledgments}
Funding from ANR-DFG (Agence Nationale de la Recherche - Deutsche Forschungsgemeinschaft) for GMT (Generalised Maxwellian Theories ANR-22-CE92-0028-01) received together with Universit\"at Bremen is acknowledged.

\bibliography{references_spallicci_080726}

\end{document}